\documentclass[reprint,prc,showpacs,superscriptaddress,preprintnumbers,amsmath,amssymb,amsfonts]{revtex4-2}
\usepackage{graphicx,latexsym,amssymb,amsmath,color,mathrsfs,booktabs,ifpdf}
\usepackage{bm}% bold math
\usepackage{epsfig,natbib}
\usepackage[T5]{fontenc}

\def\aA{$\alpha$-nucleus\ }
\def\oc{$^{16}$O+$^{12}$C\ }

\def\cc{$^{12}$C+$^{12}$C\ }
\def\oo{$^{16}$O+$^{16}$O\ }
\def\AA{nucleus-nucleus\ }

\begin{document}
\title{Systematics of supernumerary nuclear rainbow in inelastic \oc scattering}

\author{Nguyen Hoang Phuc}
\email{phuc.nguyenhoang@phenikaa-uni.edu.vn}
\affiliation{Phenikaa Institute for Advanced Study (PIAS), Phenikaa University, Hanoi 12116, Vietnam}

\author{Nguyen Tri Toan Phuc}
\affiliation{Department of Nuclear Physics, Faculty of Physics and Engineering Physics, University of Science, 
	Ho Chi Minh City 700000, Vietnam}
\affiliation{Vietnam National University, Ho Chi Minh City 700000, Vietnam}
 
\author{Do Cong Cuong} 
\affiliation{Institute for Nuclear Science and Technology, VINATOM, Hanoi 122772, Vietnam}

\date{\today}
\begin{abstract}
We perform a systematic study of inelastic nuclear rainbow scattering for the \oc system to the 2$^+$ (4.44 MeV) state of $^{12}$C at incident energies of 100--608 MeV with the coupled-channels method. The recently generalized nearside-farside decomposition for inelastic scattering was applied in combination with the multichannel deflection function analysis to elucidate the origin of the nuclear rainbow phenomenon and the suppression of the primary and supernumerary Airy minima in the inelastic scattering cross section. The systematic evolution of the Airy minima for the excited 2$^+$ (4.44 MeV) state of $^{12}$C was unambiguously determined. Our work suggests that there is no clear shift in the positions of the first Airy minima and a small shift at low energies for the second and third Airy minima between the inelastic and elastic scattering cross sections. Using the $K$-subamplitudes splitting technique combined with the generalized nearside-farside decomposition and deflection function, the distinct refractive pattern commonly suppressed in the inelastic heavy-ion scattering can be interpreted and provides new insights into the relationship between elastic and inelastic nuclear rainbow scattering. 
\end{abstract}
\maketitle

\section{Introduction}\label{sec1} 
The nuclear rainbow phenomenon in heavy-ion scattering is an intriguing subject in the study of nuclear reactions \cite{Kho07r,Bra97}. For systems exhibiting weak absorption, such as several light heavy-ion and \aA systems, incident waves can penetrate deeply into the nucleus and be refracted by the attractive nucleus-nucleus interaction. The cross section of such a process displays a distinctive nuclear rainbow pattern as a result of the interference between different components of the refracted (farside) waves. The resulting Airy structure, which consists of some deep minima followed by a broad structureless falloff, can be observed at medium and large angles of the scattering cross section. This signature structure of the nuclear rainbow is highly sensitive to the real optical potential (OP) down to small internucleus distances. This feature allows the accurate determination of the density dependence of the effective nucleon-nucleon interaction through the use of the microscopic folding model \cite{Kho07r,Bra97} as well as useful information on the cluster structure for a wide range of nuclei in a unified way \cite{Mi98,Ohk16,Ohk23,Phu18}. 

Although the calculation of the scattering process is fully quantum mechanics, the nuclear rainbow phenomenon can be more insightfully interpreted using a semiclassical approach based on either the barrier-wave and internal-wave (BI) or nearside-farside (NF) decompositions. The BI method, first proposed by Brink and Takigawa \cite{Bri77}, visualizes the scattering waves in terms of internal waves penetrating through the effective potential barrier into the nuclear interior, and barrier waves reflected at the potential barrier. On the other hand, the NF technique of Fuller \cite{Ful75,Hus84} splits the scattering amplitude into the nearside and farside components, corresponding to the waves deflected to the same (near) side and those traveling to the opposite (far) side of the scattering angle, respectively. It is now well-established that these two interpretations are complementary, and that the broad Airy oscillations found in elastic scattering cross sections at medium and large angles are the result of the interference between different components of the farside amplitude \cite{Mi00b,An01,Kho07r,Row77,Hus84,Bra97}. We also note that the NF technique has been recently extended to account for the boson symmetrization often found in strong refractive systems like \cc and \oo \cite{Phu24}.

Nuclear rainbows have also been investigated in inelastic scattering using numerous experimental data of light heavy-ion and $\alpha$-nucleus systems \cite{Kho87,Kho05,Boh82,Boh85,Boh93,Dea85,Dem87,Dem92,DAr94,DAg95,Mi04,Ohk04,Ohk07,Ham13,Ohk14i,Ohk17,Per12,Has18,Ber20}. We note that inelastic rainbow scatterings have also been reported in some atomic and molecular systems and often referred to as the potential (or $\ell$-type) and rotational rainbows (see Refs.~\cite{Hef81,Zie87,Got87,Kor92,McCa98,Cha14,Onv15,Mor21} and references therein). To study the formation mechanism of the nuclear rainbow in inelastic scattering, the BI and NF techniques in standard forms have been applied to analyze several experimental data \cite{Mi04, Dem87, Dem92, DAr94, DAg95,Dea85,Ber20,Ohk07}. The analysis results confirmed the dominance of farside scattering amplitude at large angles in the inelastic scattering cross section, but the Airy oscillation pattern is strongly suppressed and could not be clearly identified in the cross sections with nonzero transferred angular momentum (i.e. nonmonopole transitions). This damping of the refractive effect was originally suggested to be attributed to the enhanced absorption in the exit channel \cite{Kho05} or differences in the shape of the transition form factor and OP \cite{Dem92}. However, in some scattering systems, the inelastic scattering cross section magnitudes exceed those from elastic scattering at medium and large angles. This suggests the Airy refractive pattern in the inelastic scattering channel could be hindered due to other effects (see Fig.~1 of Ref.~\cite{Phu21}). 

For reaction processes involve a nonzero transferred angular momentum, the NF technique are straightforwardly generalized by using the associated Legendre functions in the nearside and farside amplitudes \cite{Ful77,Dea84,Sak90}. However, since each orbital angular momentum in entrance channel can produce several coupled counterparts in the exit channel, the total nearside and farside amplitudes can be further decomposed into different subamplitudes depends on the specific summation. Recently, a new approach for the NF decomposition technique for inelastic scattering has been proposed based on the summation of subamplitudes with the same transferred angular momentum substate $K$ \cite{Phu21}. This new NF decomposition convincingly explained the suppression of the inelastic Airy minima in terms of the smearing of the total scattering amplitude due to the coherent mixing of subamplitudes belonging to different transferred angular momentum substate $K$ groups.

Among the light heavy-ion systems known to exhibit the nuclear rainbow scattering, the \oc system is one of the most extensively investigated, with elastic and inelastic scattering data to the $2^+_1$ (4.44 MeV) state of $^{12}$C measured at various energies spanning a broad range with $E_\text{lab}$ from 20 to 1503 MeV and over an extensive angular region \cite{Rou85, Bra86,Vil89,Ogl98,Ogl00,Ogl03,Nic00,Bra01, Ohk14i,Szi06}. Several nuclear rainbow analyses of the scattering data in this energy range have been performed using the optical model (OM) \cite{Kho16, Ogl98,Ogl00,Bra88,Szi01}, coupled-channels (CC) \cite{Has18,Ohk14e,Ohk14i,Ohk17,Phu21}, and coupled-reaction-channels (CRC) \cite{Szi02,Phu18,Phu21-2,Phu21cip} approaches.

The existence of the supernumerary rainbow structure in inelastic heavy-ion scattering has been investigated using the CC analysis for the case of \oc scattering to the $2^+_1$ (4.44 MeV) state of $^{12}$C from 124 to 200 MeV \cite{Ohk17}. By observing the cross section when switching off the imaginary potential, the authors suggest a similar mechanism for the existences and properties of the supernumerary rainbow patterns between the elastic and inelastic scattering. The analysis in Ref.~\cite{Ohk17} also indicated a backward shift in scattering angles of the A1 and A2 Airy minima of the inelastic cross section compared to the elastic one. However, due to the lack of a proper splitting of farside subamplitudes as suggested in Ref.~\cite{Phu21}, the systematics of the supernumerary rainbow in inelastic scattering is difficult to determine unambiguously from the Airy-suppressed cross section \cite{Ohk17}. 

%Therefore in this study, we focus on investigating such systematics in the inelastic scattering of the \oc system using the recently developed NF decomposition.

As mentioned above, the Airy minima of the inelastic scattering cross section exciting the $2^+_1$ state of $^{12}$C are rather flat and smeared out, which can easily lead to large uncertainties in determining the Airy minima if relying solely on the experimental data or calculated cross sections without the use of additional analysis techniques. Therefore, in this study, we perform a systematic analysis of elastic and inelastic \oc scatterings over a wide energy range from 100 to 608 MeV, employing the deflection function analysis combined with the recently extended NF decomposition technique \cite{Phu21}, where inelastic scattering amplitude is split into subamplitudes having the same transferred angular momentum substate $K$. The calculations are based on the CC method with double-folding potentials and form factors. Based on the analysis results of the inelastic cross sections corresponding to these $K$-subamplitudes, we can reliably clarify the formation and energy evolution of the Airy minima of the primary and supernumerary nuclear rainbows in the \oc system. 
          
\section{Decomposition of the elastic and inelastic scattering amplitude in the CC
 framework}
\label{sec2} 
\subsection{General formalism}\label{sec2.1}
The CC equations for elastic and inelastic \AA scatterings are briefly described here. For each total angular momentum $J^\pi$ of the \AA scattering system, the CC equations are written as follows \cite{Kho00,Sat83} 
\begin{eqnarray} 
\left\{\frac{\hbar^2}{2\mu_\gamma}\left[\frac{d^2}{dR^2}+k^2-
\frac{L(L+1)}{R^2}\right]-\langle\gamma(LI)J |V|\gamma(LI)J\rangle \right\}
\ \ \nonumber \\
\times\chi_{\gamma J}(k,R)=\sum_{\gamma' L'I'}\langle \gamma(LI)J|V|\gamma'(L'I')J\rangle 
\chi_{\gamma' J}(k',R),\ \ \ \label{eq1}
\end{eqnarray}
where $\hbar k=\sqrt{2\mu_\gamma E_\gamma}$ is the center-of-mass (c.m.) momentum, $\mu_\gamma$ is the reduced mass, and $\chi_{\gamma J} (k,R)$ is the scattering wave function at the internuclear distance $R$. The entrance and exit channels are indicated by $\gamma$ and $\gamma'$, respectively. The orbital angular momenta $L$ (entrance channel) and $L'$ (exit channel) are determined according to the angular momentum conservation $\bm{J}=\bm{L}+\bm{I}=\bm{L'}+\bm{I'}$ with the spin $I$ ($I'$) being the spin of the entrance (exit) channel. In this work, we consider the incident nuclei in their spinless ground states ($I=0$) and the target nucleus in the exit channel being excited to a state with spin $I'$. The diagonal matrix element $V_{\gamma\gamma}(R)\equiv \langle\gamma(LI)J |V|\gamma(LI)J\rangle$ of the projectile-target interaction is the OP, while the off-diagonal matrix element $V_{\gamma\gamma'}(R)\equiv \langle \gamma(LI)J|V|\gamma'(L'I')J\rangle$ is the transition potential, also known as the inelastic scattering form factor. Solving the CC equations (\ref{eq1}) gives the elastic scattering amplitude
\begin{equation}
 f(\theta)=f_{\rm C}(\theta)+\frac{1}{2ik}\sum_L(2L+1)\exp(2i\sigma_L)
 (S_L-1)P_L\left(\cos\theta\right), \label{eq2} 
\end{equation}
where $f_{\rm C}(\theta)$ is the Coulomb scattering amplitude with the Coulomb phase shift $\sigma_L$. $S_L$ is the diagonal element of the elastic scattering $S$ matrix, and $P_L(\cos\theta)\equiv P_{L}^{0}(\cos\theta)$ is the Legendre function of the first type. The amplitude of the inelastic scattering to an excited state of the target with spin $I'$ and projection $M_{I'}$ is explicitly expressed in the CC framework \cite{Sat83} as
\begin{eqnarray}
f_{M_{I'}}(\theta,\phi)&=&\frac{\sqrt{4\pi}}{2ik}\sum_{LL'}\sqrt{2L+1}
\langle L'-M_{I'} I' M_{I'}|L 0 \rangle \nonumber \\ 
&& \times \exp[i(\sigma_L+\sigma'_{L'})] S_{L'L} Y_{L'}^{-M_{I'}}(\theta,\phi). \label{eq3}
\end{eqnarray}
Here $Y_{LM}(\theta,\phi)$ is the spherical harmonics, $S_{L'L}$ is the matrix element of the inelastic scattering $S$ matrix, and the Coulomb phase shift $\sigma'_{L'}$ is calculated from the c.m. momentum $k'$ in the exit channel. The orbital angular momenta $L'$ in the exit channel are determined through the orbital angular momenta $L$ in the entrance channel and the excited target's spin $I'$ according to the triangle rule
\begin{equation}
 L'=L-I',L-I'+2,...,L+I'-2,L+I', \label{eq4}
\end{equation}
in which parity conservation requires a change of two units in angular momentum. We can observe that, for each value of angular momentum $L$, there are multiple coupled angular momenta $L'$ contributing to the total inelastic scattering amplitude $f_{M_{I'}}$ according to expression (\ref{eq4}).

\subsection{Nearside-farside decomposition of the scattering amplitudes}

The decomposition of the elastic scattering amplitude into nearside and farside components, using Fuller’s method \cite{Ful75}, is very helpful for illustrating the refractive Airy structure of the nuclear rainbow. Specifically, by breaking down the Legendre function $P_L(\cos \theta )$ into waves that move in opposite directions around the scattering center, the elastic scattering amplitude $f(\theta)$ can be represented in terms of nearside ($f^{\rm N}(\theta)$) and farside ($f^{\rm F}(\theta)$) components as
\begin{eqnarray} 
f(\theta)&=&f^{\rm N}(\theta)+f^{\rm F}(\theta)=f^{\rm N}_{\rm C}(\theta)+
 f^{\rm F}_{\rm C}(\theta) \nonumber \\ && +\frac{1}{2ik}\sum_L(2L+1)\exp(2i\sigma_L)(S_L-1)\nonumber \\ 
&&\times \left[\tilde Q_L^{(-)}(\cos\theta)+\tilde Q_L^{(+)}(\cos\theta)\right], 
\label{eq5}
\end{eqnarray} 
where $f^{\rm N(F)}_{\rm C}(\theta)$ is the nearside (farside) component of the Coulomb amplitude \cite{Ful75}, $\tilde Q_L^{(-)}(\cos\theta)$ and $\tilde Q_L^{(+)}(\cos\theta)$ represent the components of the nearside and farside amplitudes, respectively.  
\begin{equation} 
 \tilde Q_L^{(\mp)}(\cos\theta)=\frac{1}{2} 
 \left[P_L(\cos\theta)\pm {2i\over\pi}Q_L(\cos\theta)\right]. \label{eq6}
\end{equation}
The Legendre function of the second kind is denoted by $Q_L(\cos\theta)\equiv Q_{L}^{0}(\cos\theta)$. It is widely recognized \cite{Bra97,Kho07r,Hus84} that the farside component of the elastic amplitude (\ref{eq5}) governs the nuclear rainbow pattern observed in elastic \aA and light HI scattering. This nuclear rainbow pattern manifests as broad oscillations of the Airy minima at medium and large scattering angles.

It is straightforward to extend the NF decomposition to inelastic scattering. For each projection $M_{I'}$ of the target spin $I'$, the NF decomposition (\ref{eq5}) is generalized to decompose the inelastic scattering amplitude (\ref{eq3}) using the associated Legendre functions as 
\begin{eqnarray}  
 f_{M_{I'}}(\theta,\phi)&=&f^{\rm N}_{M_{I'}}(\theta,\phi)+
   f^{\rm F}_{M_{I'}}(\theta,\phi) \nonumber\\
   &=&\frac{\sqrt{4\pi}}{2ik}\sum_{LL'}\sqrt{2L+1}
\langle L'-M_{I'}I'M_{I'}|L 0 \rangle A_{L'}^{M_{I'}} \nonumber\\
&& \times  \exp[i(\sigma_L+\sigma'_{L'})]\exp(-iM_{I'}\phi)S_{L'L}  \nonumber\\
&& \times   \left[\tilde Q_{L'}^{-M_{I'}(-)}(\cos\theta)+
\tilde Q_{L'}^{-M_{I'}(+)}(\cos\theta)\right], \nonumber \\ \label{eq7}
\end{eqnarray}
where
\begin{equation} 
\tilde Q_{L}^{M(\mp)}(\cos\theta)={1\over 2}
\left[P_{L}^{M}(\cos\theta)\pm {2i\over\pi}Q_{L}^{M}(\cos\theta)\right],
\end{equation} 
and  
\begin{equation} 
 A_{L}^M=\sqrt{\frac{2L+1}{4\pi}\frac{(L+M)!}{(L-M)!}}.
\end{equation} 
$P_{L}^M(\cos\theta)$ and $Q_{L}^M(\cos\theta)$ denote the associated Legendre functions of the first and second kinds, respectively.
\subsection{Mixing of K-subamplitudes} 
\label{sec2.2}
From Eq.~(\ref{eq3}), it is evident that, for $I'\neq 0$, coupled partial waves of different substates ($\bm{z}$-components) of transferred angular momentum can contribute to the inelastic scattering amplitude at a given scattering angle $\theta$. Expressing the selection rule (\ref{eq4}) as
\begin{equation}
	L'=L+K; \quad K=-I', -I' + 2, \ldots, I' - 2, I', \label{eq:sel}
\end{equation}
the inelastic scattering amplitude (\ref{eq3}) can be represented in terms of the $K$-subamplitudes as
\begin{equation}
 f_{M_{I'}}(\theta,\phi)=\sum_{K=-I'}^{I'} f^{K}_{M_{I'}}(\theta,\phi). \label{eq8}
\end{equation} 
Similar to the elastic amplitude (\ref{eq2}), each $K$-subamplitude of $f_{M_{I'}}(\theta,\phi)$ can be expressed in terms of the orbital momenta of the entrance channel $L$ as follows
\begin{eqnarray}
 f^{K}_{M_{I'}}(\theta,\phi)&=&\frac{\sqrt{4\pi}}{2ik}\sum_{L}\sqrt{2L+1}
 \langle (L+K)-M_{I'} I' M_{I'}|L 0 \rangle \nonumber \\ 
 & & \times\exp[i(\sigma_L+\sigma'_{L+K})]S_{(L+K)L} Y_{(L+K)}^{-M_{I'}}(\theta,\phi). \nonumber \\
 \label{eq9} 
\end{eqnarray} 
The contribution from each $K$-subamplitude at the given scattering angle to the partial inelastic scattering cross section is defined as 
\begin{equation}
 \frac{d\sigma_K}{d\Omega}=\sum_{M_{I'}}
 \left|f^{K}_{M_{I'}}(\theta,\phi)\right|^2, \label{eq10}
\end{equation} 
so the coherently summed cross section of the inelastic scattering to a target excited state with spin $I'$ is given by
\begin{equation}
\frac{d\sigma_{\rm inel}}{d\Omega}=\sum_{M_{I'}}
\left|\sum_{K=-I'}^{I'} f^{K}_{M_{I'}}(\theta,\phi)\right|^2.   \label{eq11}
\end{equation} 
Therefore, the inelastic scattering cross section for an excited state with spin $I' \neq 0$ is described by the interference of the $K$-subamplitudes, where $K$ takes values from $-I'$ to $I'$ as in (\ref{eq:sel}). Each $K$-subamplitude is summed over the orbital angular momentum $L$ similar to the elastic scattering amplitude. Note that the azimuthal angle $\phi$ in (\ref{eq9}) has been canceled out in Eqs.~(\ref{eq10}) and (\ref{eq11}) due to the modulus square.  

Similarly to Eq.~(\ref{eq7}), we obtain the nearside and farside components of each inelastic scattering $K$-subamplitude (\ref{eq9}) as
\begin{eqnarray} 
 f^{K}_{M_{I'}}(\theta,\phi)&=&f^{K, {\rm N}}_{M_{I'}}(\theta,\phi)+
   f^{K,{\rm F}}_{M_{I'}}(\theta,\phi)\nonumber \\ 
&=&\frac{\sqrt{4\pi}}{2ik}\sum_L\sqrt{2L+1}
\langle (L+K)-M_{I'}I'M_{I'}|L 0\rangle\nonumber \\ 
&&\times A_{(L+K)}^{M_{I'}}\exp[i(\sigma_L+\sigma'_{L+K}-M_{I'}\phi)] \nonumber \\ 
&&\times \left[\tilde Q_{(L+K)}^{-M_{I'}(-)}(\cos\theta)+
\tilde Q_{(L+K)}^{-M_{I'}(+)}(\cos\theta)\right] \nonumber \\
&& \times S_{(L+K)L}. \label{eq12}
\end{eqnarray}   
This generalized NF decomposition (\ref{eq12}) allows us to identify the nearside and farside contributions from each $K$-subamplitude (\ref{eq9}), in a similar manner with the elastic scattering NF decomposition, to further explore the mechanism behind the nuclear rainbow formation in inelastic \AA scattering. 
\subsection{Deflection functions} 
As mentioned above, the nuclear rainbow structure in elastic scattering is dominated by the contribution of the farside component, with the Airy oscillation pattern observed at large angles originating from the interference between the $f^{\rm F_<}$ and $f^{\rm F_>}$ components of the farside amplitude \cite{Hus84,Bra97,An01,Phu24}. These two components $f^{\rm F_<}$ and $f^{\rm F_>}$ are divided at $L=L_{\rm R}$ into two domains corresponding to $L_<$ and $L_>$, where $L_{\rm R}$ is the angular momentum associated with the rainbow angle $\Theta_{\rm R}$ \cite{Hus84}. Here, $\Theta_{\rm R}\equiv\Theta(L_{\rm R})$ corresponds to the minimum of the deflection function $\Theta$, defined as follows \cite{Kho07r,Bra97}
\begin{equation}
\Theta(L)=2\frac{d(\delta_L+\sigma_L)}{dL},
\end{equation}
where $\delta_L$ represents the real nuclear phase shift of elastic scattering channel with the imaginary part are turned off. Thus, the appearance of a minimum in the deflection function is the basis for the formation of a nuclear rainbow \cite{Bra97,Hus84}.

In analogy to the above mechanism of nuclear rainbow formation in elastic scattering, a similar rainbow formation mechanism is expected to occur in inelastic scattering. The farside component of the inelastic scattering amplitude for each $K$ component in Eq.~(\ref{eq12}) can also be extended and divided into two components $f^{(K),\rm F<}$ and $f^{(K),\rm F>}$ at $L'=L'_{\rm R}$, where $L'_{\rm R}$ is the angular momentum associated with the rainbow angle $\Theta^K_{\rm R}$. Here, $\Theta^K_{\rm R}$ is the minimum of the inelastic deflection function corresponding to the subamplitudes for each $K$ in Eq.~(\ref{eq9}), which we introduced as
\begin{equation}
\Theta^K(L)= \frac{d\left[2\delta_{(L+K)L}+\sigma_L+\sigma'_{L+K}\right]}{dL},
\end{equation}
where $\delta_{(L+K)L}$ represents the real inelastic scattering phase shift. In addition to examining the partial inelastic scattering cross sections for each $K$, investigating the deflection functions corresponding to these component amplitudes provides further insights into the formation of nuclear rainbows in inelastic scattering. It is noted that although the deflection function analyses have been carried out for inelastic scatterings of atomic and molecule systems \cite{Mor21,Cha14,Onv15}, such investigations have not yet been done for nuclear inelastic scatterings. 
\section{Refractive elastic and inelastic \oc scattering}

\begin{table*}[bht]
	\begin{center}
		\caption{The parameters of the OPs (\ref{eq13}) used in the CC calculations (\ref{eq1}) of the 
			elastic and inelastic \oc scatterings. $J_{ \rm R}$ and $J^{\rm en(ex)}_{\rm W}$ are the volume 
			integrals per interacting nucleon pair of the real and imaginary parts of the OP, 
			respectively.} \label{t1} 
		\vspace*{0cm}\hspace*{0cm}
		\begin{tabular}{ccccccccc} 
			\hline\hline\
			$E_{\rm lab}$ & $N_{\rm R}$ & $J_{\rm R}$ & $W^{\rm en}_{V}$ & $W^{ \rm ex}_{V}$ & $R_{ V}$ & $a_{V}$ & $J^{\rm en}_{\rm W}$ & $J^{ \rm ex}_{\rm W}$ \\
			(MeV) &  & (MeV~fm$^3$) & (MeV) & (MeV) & (fm) & (fm) & (MeV~fm$^3$) & (MeV~fm$^3$) \\ \hline
			100   & 0.920 & 303.7 & 10.04  &	13.00 & 5.600 &	0.500 & 41.51	& 53.74 \\ 
			115.9 & 0.925 & 303.5 & 9.705  &    13.74 & 5.647 &	0.549 & 41.68   & 59.00 \\ 	
			124   & 0.924 & 302.3 & 11.41  &	14.50 &	5.650 &	0.542 & 49.00   & 62.24 \\	
			170   & 0.921 & 296.1 & 13.50  &	17.50 &	5.897 &	0.609 & 66.73	& 86.50 \\
			181   & 0.927 & 296.7 & 13.96  &	18.00 &	5.781 &	0.650 & 66.19   & 85.34 \\	
			200   & 0.926 & 294.3 & 13.50   &	17.99 &	6.100 &	0.520 & 71.65	& 95.45 \\	
			260   & 0.925 & 287.3 & 18.00   &	18.50 &	5.800 &	0.550 & 83.42	& 85.74 \\	
			281   & 0.930 & 286.6 & 18.50   &	19.50 &	5.756 &	0.550 & 83.92	& 88.46 \\	
			608   & 0.915 & 247.7 & 19.30   &	19.30 &	5.745 &	0.586 & 88.02	& 88.02 \\	
			\hline\hline 
		\end{tabular}
	\end{center}
\end{table*} 

\begin{figure}[bt]\vspace*{0.1cm}\hspace*{-0.4cm}
	\includegraphics[angle=0,width=0.53\textwidth]{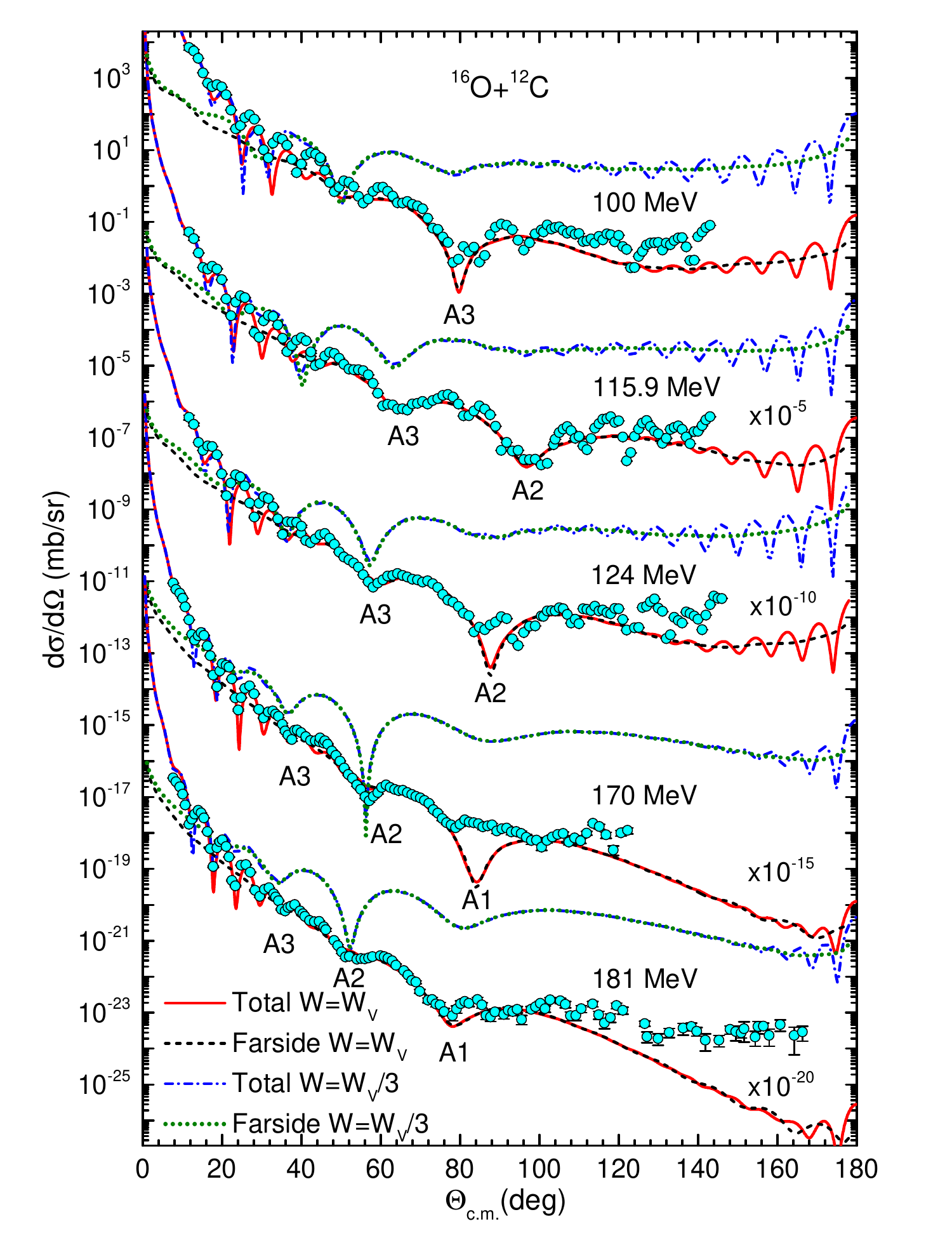}\vspace*{-0.2cm}
	\caption{CC description (solid and dashed-dotted lines) of the elastic \oc scattering data measured at $E_{\rm lab} = 100$--$181$ MeV \cite{Ogl00,Ogl03,Nic00} with different absorptive strengths $W^{\rm en(ex)}_V$ of the WS imaginary OP. The farside cross sections (dashed and dotted lines) are given by the NF decomposition (\ref{eq5}) . A$k$ is the $k$-th order Airy minimum.} \label{f1}
\end{figure}

\begin{figure}[bt]\vspace*{0cm}\hspace*{-0.2cm}
	\includegraphics[angle=0,width=0.51\textwidth]{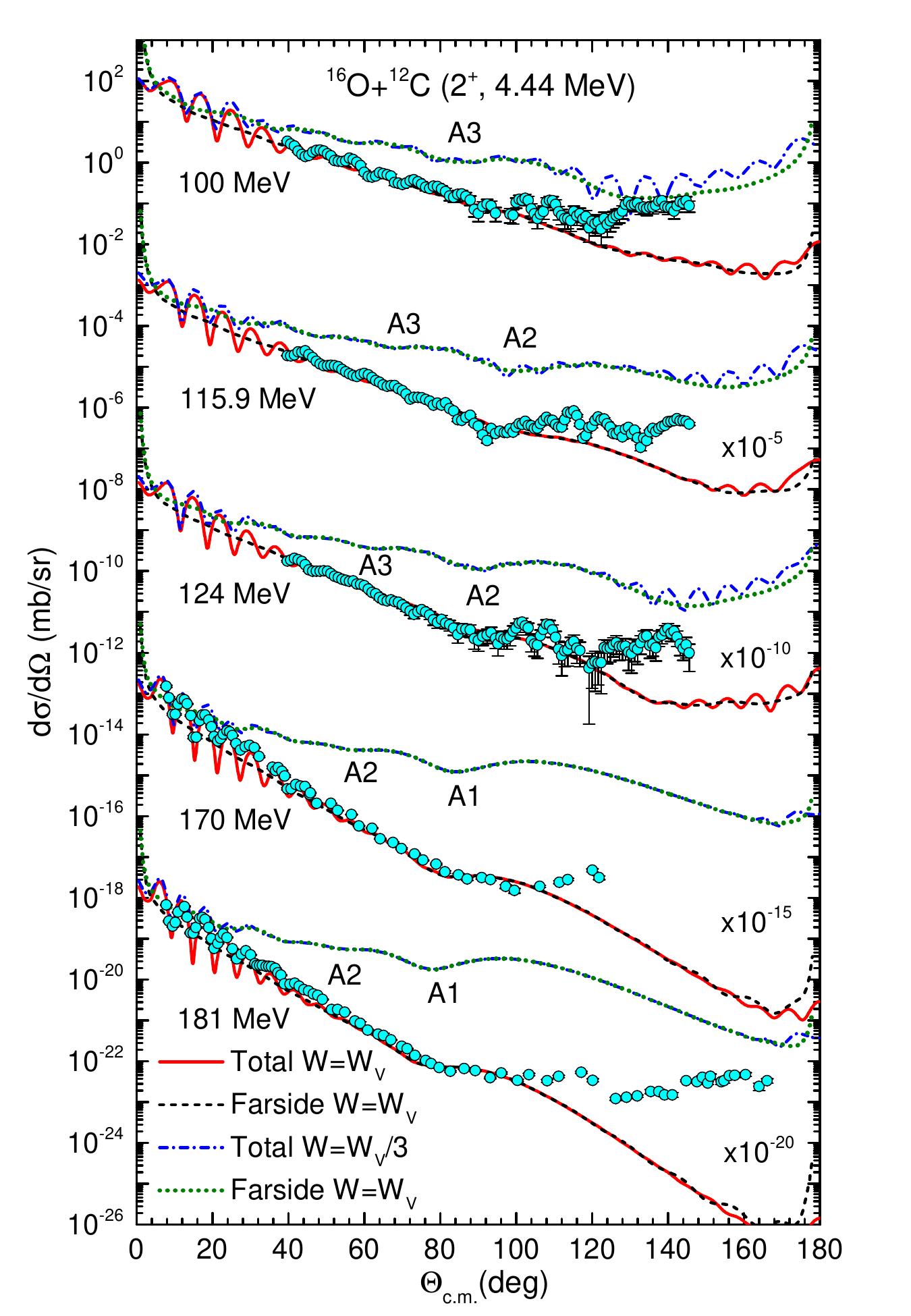}\vspace*{-0.2cm}
	\caption{CC description (solid and dashed-dotted lines) of the inelastic \oc scattering data to the $2^+_1$ state of $^{12}$C measured at $E_{\rm lab} =
		100$--$181$ MeV \cite{Ohk14i,Szi06} with the same real OP but with different
		absorptive strengths $W^{\rm en(ex)}_V$ of the WS imaginary OP. The farside cross sections (dashed and dotted line) are given by
		the NF decomposition (\ref{eq7}) . A$k$ is the $k$-th order Airy minimum.} \label{f2}
\end{figure}

To calculate the elastic and inelastic scattering cross sections, the matrix elements of the nucleus-nucleus interaction in the CC equations (\ref{eq1}) need to be determined. The diagonal matrix elements $V_{\gamma\gamma(\gamma'\gamma')}(R)$ of the projectile-target interaction are determined by the total OPs $U_{\rm en(ex)}(R)$ \cite{Kho00} of the entrance and excited channels. The double-folding model \cite{Kho00} is employed to calculate the real part of the OP using the latest version of the density-dependent CDM3Y3 interaction with the rearrangement term taken into account \cite{Kho16}. For the nuclear ground-state densities, we adopt the Fermi distribution, with the two parameters chosen to reproduce the empirical matter radii of $^{16}$O and $^{12}$C nuclei \cite{Kho01}. Since the nuclear rainbow structure is also affected by the absorption of the interaction, which is difficult to obtain accurately from the microscopic calculations, the imaginary part of the OP is usually chosen in a flexible Woods–Saxon form. Thus, we have
\begin{eqnarray}
 U_{\rm en(ex)}(R)=N_{\rm R}V(R)+ iW_{ \rm en(ex)}(R)+V_{\rm C}(R), \label{eq13} 
\end{eqnarray}
where
\begin{eqnarray}
W_{\rm en(ex)}(R)=-\frac{W^{\rm en(ex)}_V }{1+\exp[(R-R_V)/a_V]}. \label{eq13b} 
\end{eqnarray}
The Coulomb potential $V_{\rm C}(R)$ is generated by folding two uniform charge distributions, with their mean-squared radii chosen to reproduce measured charge radii of the two nuclei. Here, the real part of the OP in the exit channel is assumed to be equal to the one in the entrance channel, with $N_{\rm R}$ being the renormalization factor for the real folding potential. We note that in Ref.~\cite{Ohk17}, the folding calculations using microscopic densities for the $^{12}$C nucleus also showed that the potentials for the entrance channel and the $2^+_1$ exit channel are almost the same (see Fig.~4 of Ref.~\cite{Ohk17}). The difference between the OPs in the entrance and exit channels in our work lies in the strength of the imaginary potentials $W_{\rm en}(R)$ and $W_{\rm ex}(R)$, whose parameters are adjusted to achieve the best CC descriptions of the elastic and inelastic scattering data. 

The off-diagonal matrix element $V_{\gamma\gamma'}(R)$ in the CC equation (\ref{eq1}) is the transition potential $U_{I'}(R)$ from the elastic channel to the excited one with the target spin $I'$, determined as ~\cite{Kho00}
\begin{equation}
  U_{I'}(R)=N_{\rm R}V_{I'}(R)-i\delta_{I'}\frac{\partial W_{\rm en}(R)}
 {\partial R}+ V_{{\rm C},I'}(R), \label{eq15} 
\end{equation}
where $V_{I'}(R)$ and $V_{{\rm C},I'}(R)$ are the inelastic form factors for the real nuclear potential and the Coulomb potential, respectively. These inelastic form factors are obtained using the double-folding model with the transition density of $^{12}$C given by the resonating group method (RGM) \cite{Kam81}. Meanwhile, the imaginary form factor is obtained with the nuclear deformation lengths $\delta_{I'}$ given by the collective-model prescription \cite{Kho00} using the measured $B(EI')$ transition rates of the considered excited states of $^{12}$C \cite{Ram87,Kib02}. The CC calculations are performed using the ECIS code of Raynal \cite{Ray72}, which provides the $S_L$ and $S_{LL'}$ matrix elements used for the $K$-subamplitudes and NF analysis according to the decomposition (\ref{eq5}), (\ref{eq7}), (\ref{eq9}), and (\ref{eq12}). In this analysis, we do not consider the reorientation effect, as our study primarily focuses on the angular region of the three A1, A2, and A3 Airy minima. As shown in Ref.~\cite{Ohk15}, the reorientation effect mostly affect the angular region following the broad shoulder of the primary A1 minimum. Therefore, it is expected that the reorientation effect does not influence the systematics investigated in this work.

\begin{figure}[bt]\vspace*{0cm}\hspace*{-0.4cm}
	\includegraphics[angle=0,width=0.53\textwidth]{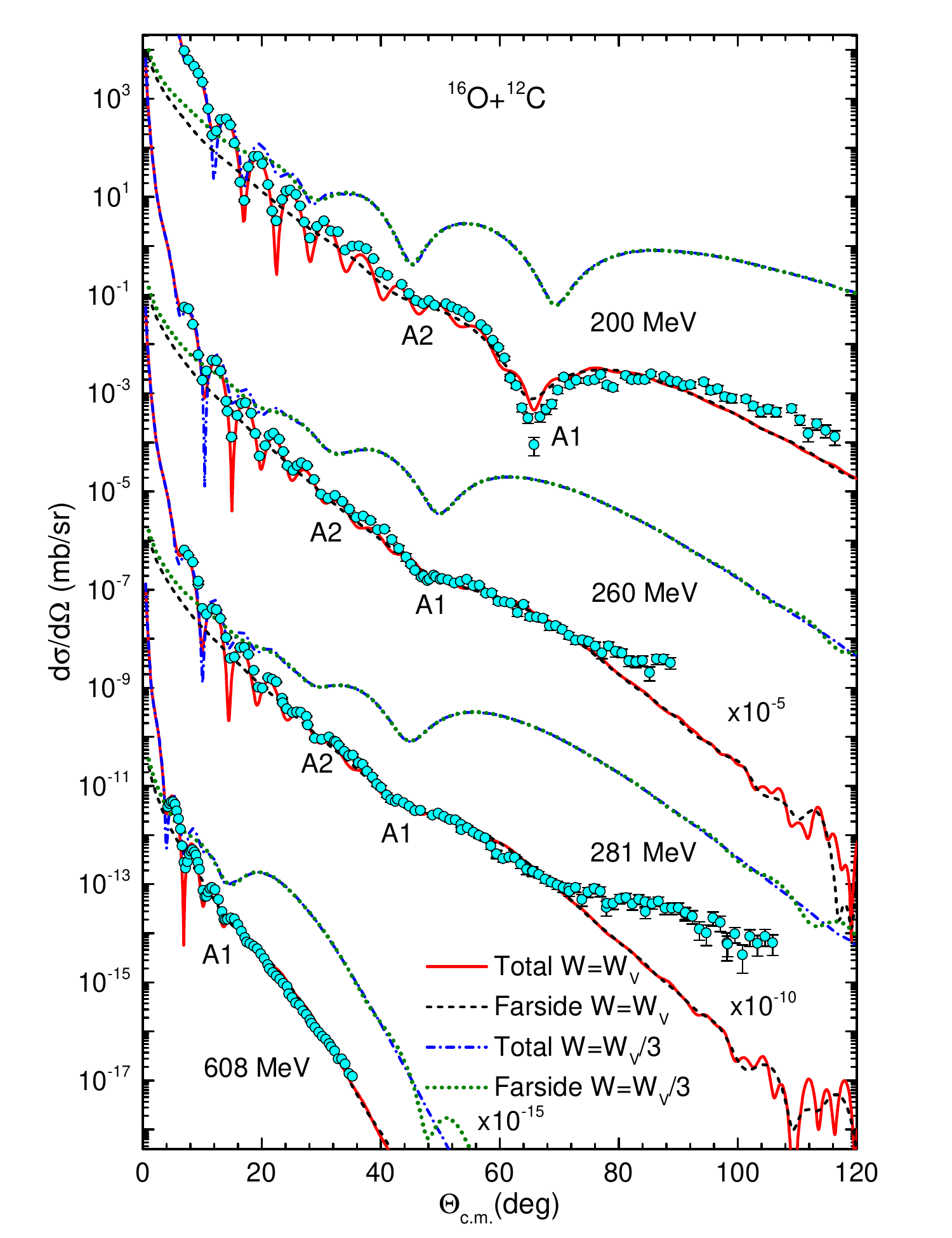}\vspace*{-0.2cm}
	\caption{Same as Fig.~\ref{f1} but for the elastic \oc scattering
data measured at $E_{\rm lab} = 200$--$608$ MeV \cite{Ogl00,Ogl03,Bra86}.} \label{f3}
\end{figure}

\begin{figure}[bt]\vspace*{0.1cm}\hspace*{-0.2cm}
	\includegraphics[angle=0,width=0.51\textwidth]{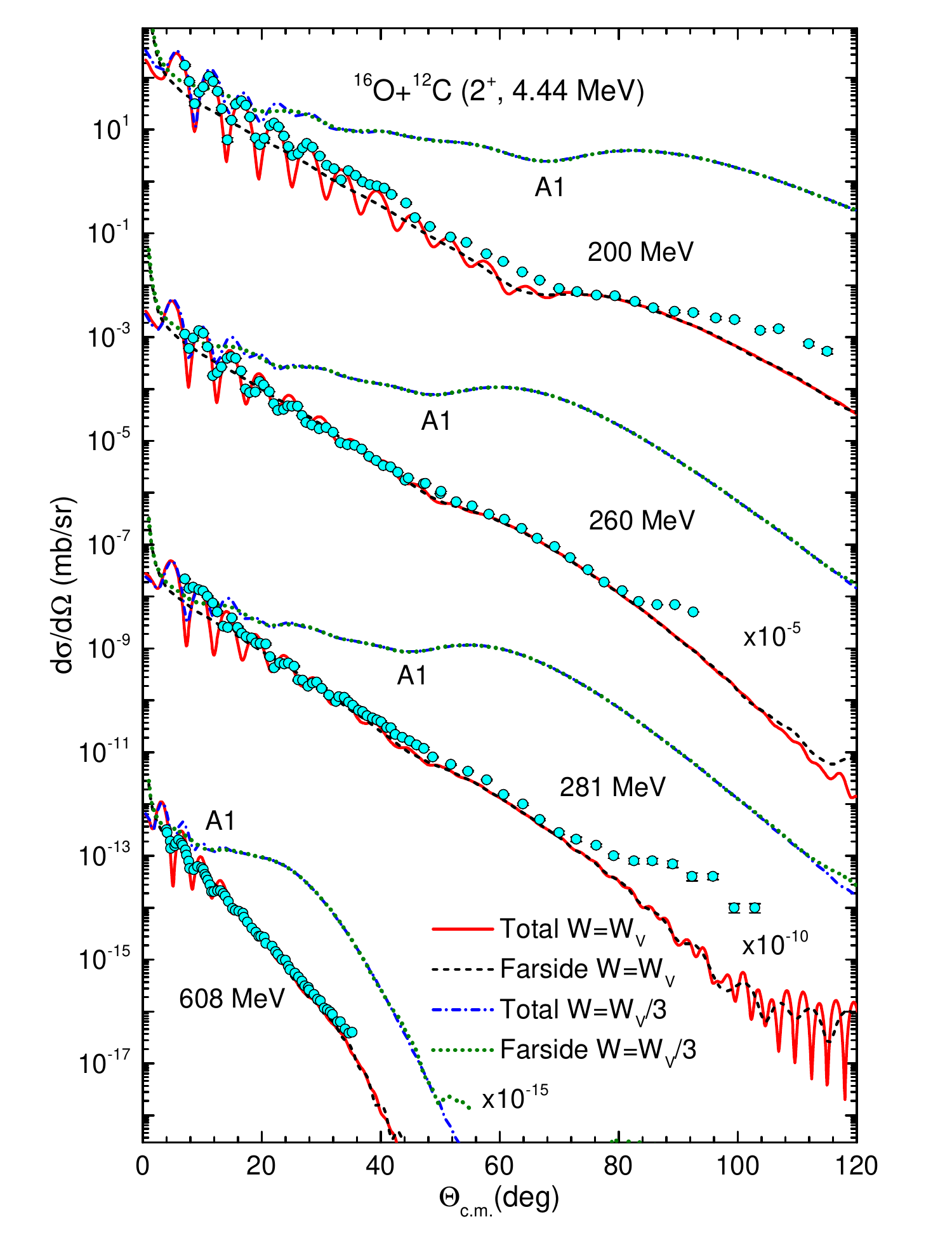}\vspace*{-0.2cm}
	\caption{Same as Fig.~\ref{f2} but for the inelastic  $^{12}$C of \oc scattering
		data to the $2^+_1$ state measured at $E_{\rm lab} = 200$--$608$ MeV \cite{Ohk14i,Bra86}.} \label{f4}
\end{figure}

In this work, we analyze the cross-section data for the elastic \cite{Nic00,Ogl00,Ogl03,Bra86} and inelastic scatterings to the $2^+_1$ state of $^{12}$C \cite{Szi06,Bra86,Ohk14i} at nine incident energies of 100--608 MeV. The analysis in this work covers the complete refractive energy region of the \oc system for both the elastic and inelastic scatterings. Figures \ref{f1}--\ref{f4} illustrate the CC calculation results for the elastic and inelastic scatterings cross-section data, with the parameters of the imaginary part of the OP and the renormalization factor $N_{\rm R}$ for the real part presented in Table~\ref{t1}. To simultaneously describe the experimental data for elastic and inelastic scatterings \cite{Ohk14i,Szi06,Ogl00,Ogl03,Nic00,Bra86}, the strength of the imaginary potential in the exit channel is required to be larger than that in the entrance channel (see Table~\ref{t1}). A similar practice has also been employed previously to describe the inelastic \oo scattering \cite{Kho05}. The calculation results provide a good description of the experimental data at angles up to the refractive region. For backward angular regions, a complete description required numerous couplings to other inelastic and transfer channels \cite{Phu18,Ohk14e}. Since this work only focuses on the angular region where the Airy structure clearly manifests without strong disturbances from higher order effects, we do not consider channels other than the ground and first excited states of $^{12}$C.     

In the energy region displayed in Figs. \ref{f1} and \ref{f3}, most of the primary and supernumerary nuclear rainbow structures of elastic scattering can be clearly observed experimentally. At the lower energies $E_{\rm lab}=100, 115.9$, and 124 MeV, the first Airy minimum A1 remains hidden in the backward angular region, and only the second and third Airy minima A2 and A3 are visible, with $\theta_{\rm c.m}\approx80^\circ$ (A3) at $E_{\rm lab}=100$ MeV, $\theta_{\rm c.m}\approx 97^\circ$ (A2) and $\theta_{\rm c.m.}\approx 64^\circ$ (A3) at $E_{\rm lab}=115.9$ MeV, and $\theta_{\rm c.m.}\approx 88.5^\circ$ (A2) and $\theta_{\rm c.m.}\approx 60^\circ$ (A3) at $E_{\rm lab}=124$ MeV. These values are consistent with previous studies of \oc elastic scattering at lower energies between 85 MeV and 132 MeV \cite{Szi01,Phu18}, and will be later investigated as part of the nuclear rainbow systematics below.
 
At higher energies, the first Airy minimum of elastic scattering cross section shifts towards the small angular region, and thus the A1, A2, and A3 minima can be identified with values determined as $\theta_{\rm c.m.}\approx 84^\circ$, $\theta_{\rm c.m.} \approx 58^\circ$, and $\theta_{\rm c.m.}\approx 37^\circ$ at $E_{\rm lab}=170$ MeV, and $\theta_{\rm c.m.} \approx 78^\circ$, $\theta_{\rm c.m.} \approx 52^\circ$, and $\theta_{\rm c.m.}\approx 35^\circ$ at $E_{\rm lab}=181$ MeV, respectively. For the case at 170 MeV, the experimental A1 value of $\theta_{\rm c.m.} \approx 79^\circ$ on the elastic scattering cross section is slightly shifted compared to the OM analysis result of $\theta_{\rm c.m.} \approx 84^\circ$. This is attributed to the elastic ($Q=0$) alpha transfer process $^{12}$C( $^{16}$O, $^{12}$C) $^{16}$O, which enhances the measured elastic cross section at backward angles and perturbs the A1 Airy position \cite{Ogl00}. 

When the projectile energy increases to $E_{\rm lab}=200$ and 260 MeV, the A1 and A2 Airy minima shift further towards forward angles, and the nuclear rainbow structure is no longer affected by the alpha transfer process. The positions of A1 and A2 can be clearly determined as $\theta_{\rm c.m.}\approx 66^\circ$ and $\theta_{\rm c.m.}\approx 47^\circ$ at $E_{\rm lab}=200$ MeV, and $\theta_{\rm c.m.}\approx 49^\circ$ and $\theta_{\rm c.m.}\approx 31^\circ$ at $E_{\rm lab}=260$ MeV, respectively. At $E_{\rm lab}=281$ and 608 MeV, the Airy minimum A2 shifts to the diffractive (Fraunhofer) part of the elastic cross section and is no longer visible in the measured data. The Airy minimum A1 at these two energies is shallow and can be more clearly identified with a weaker absorptive strength of the imaginary OP ($W_V^{\rm en(ex)}\rightarrow W_V^{\rm en(ex)}/3$), with $\theta_{\rm c.m.}\approx 45^\circ$ and $\theta_{\rm c.m.} \approx 14^\circ$ at 281 and 608 MeV, respectively.

In contrast to the elastic scattering cross section, both the calculated results and experimental data in Figs. \ref{f2} and \ref{f4} show that the Airy minima of the inelastic cross section exciting to the $2^+_1$ state of $^{12}$C are very shallow, although the NF analysis clearly demonstrates the dominance of the refractive effect associated with the farside component in the cross section at medium angles. The nuclear rainbow structure in the inelastic scattering cross section can only be faintly observed based on the calculation results when the absorption is reduced (in our case, by setting $W_V^{\rm en(ex)}\rightarrow W_V^{\rm en(ex)}/3$). This result is similar to the recent study by Ohkubo \emph{et al.} \cite{Ohk17}, where the primary and supernumerary rainbow structures of the inelastic scattering cross section exciting the $2^+_1$ state of $^{12}$C were also determined based on the technique of switching off the imaginary potentials in the $2^+_1$ channel (see Figs. 1 and 2 in Ref.~\cite{Ohk17}). Notably, even with the reduced imaginary part of the OP, the Airy minimum A2 can only be subtly seen at energies lower than 181 MeV. When the projectile energy increases to $E_{\rm lab}=200$ MeV, the supernumerary rainbow structure, starting from the A2 minimum, is no longer presented in the inelastic scattering cross section, while in the elastic cross section, one can identify the clear supernumerary structure of the nuclear rainbow, including the first three (A1, A2, A3) Airy minima.

\begin{figure}[bt]\vspace*{0cm}\hspace*{-0.5cm}
	\includegraphics[angle=0,width=0.55\textwidth]{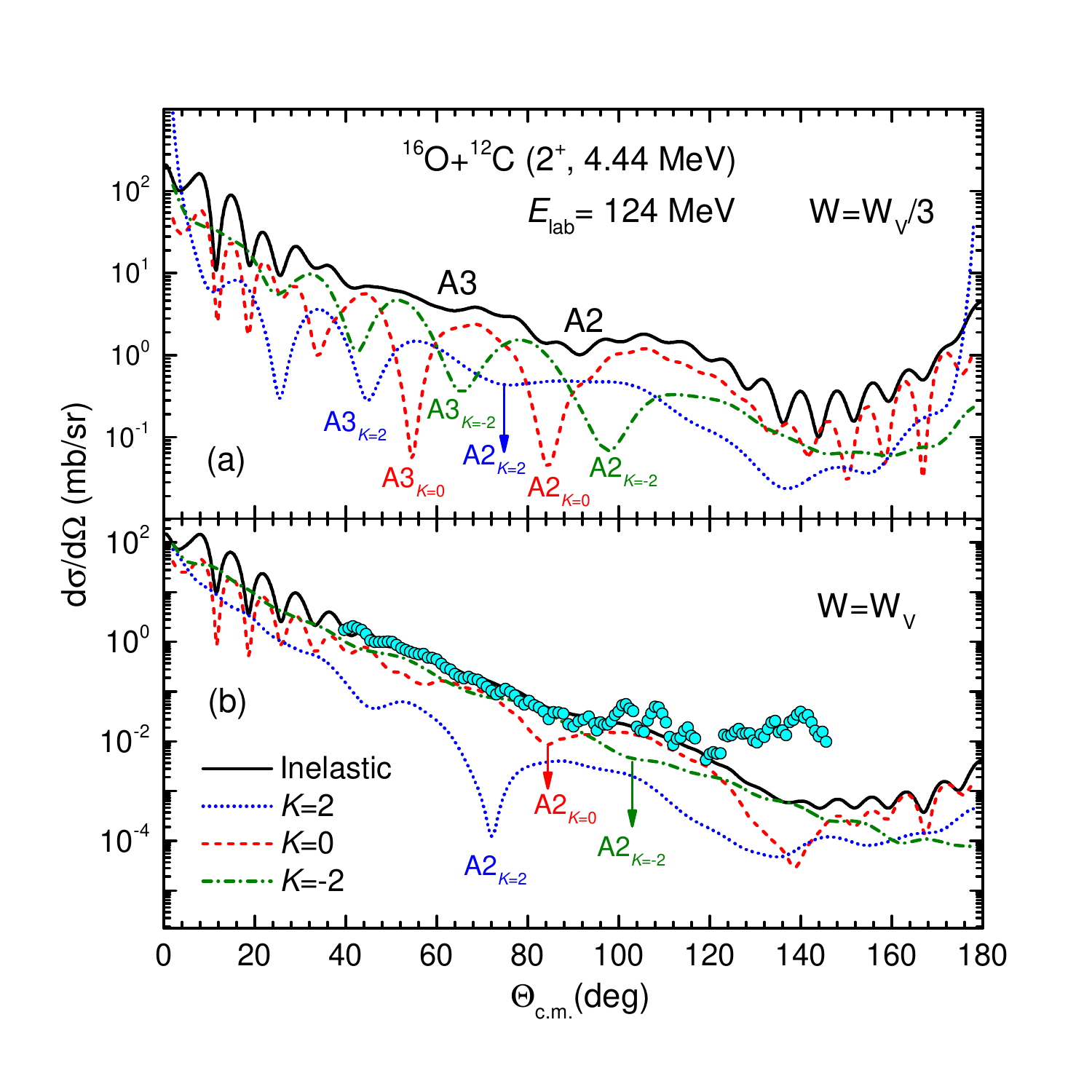}\vspace*{-0.8cm}
	\caption{CC description (solid lines) of the inelastic \oc scattering to the $2^+_1$
state of $^{12}$C  at $E_{\rm lab}$ = 124  MeV \cite{Szi06} with the same real OP but with different
absorptive strengths $W^{\rm en(ex)}_V$ of the WS imaginary OP. The dotted, dashed and dashed-
dotted-dotted lines are the partial inelastic cross sections (\ref{eq9})-(\ref{eq10}) given by the
subamplitudes with $K = 2, 0,$ and $-2$, respectively. The prominent Airy
minimum A2 is shown explicitly for each partial inelastic cross section} \label{f5}
\end{figure}

\begin{figure}[bt]\vspace*{0cm}\hspace*{-0.5cm}
	\includegraphics[angle=0,width=0.55\textwidth]{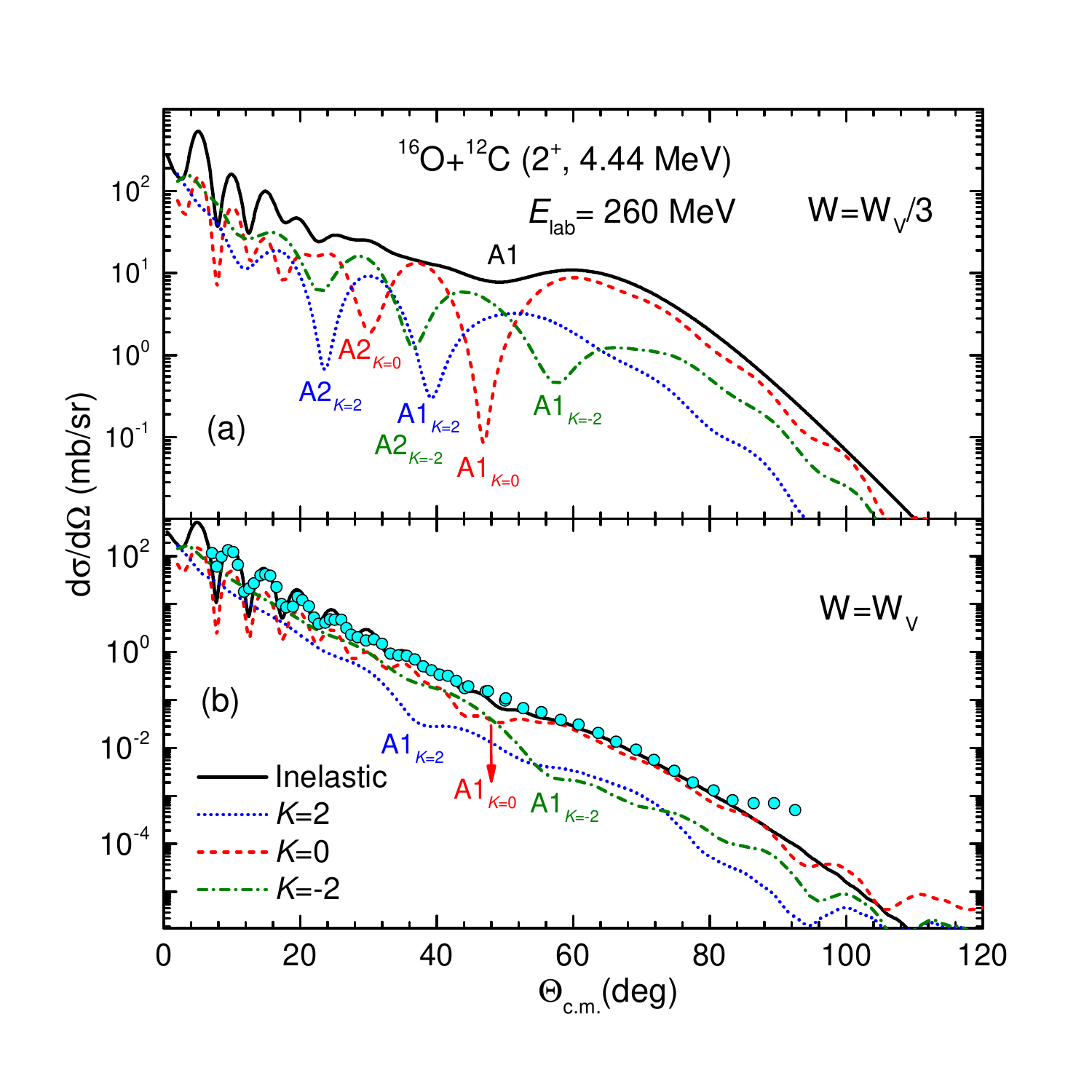}\vspace*{-0.8cm}
	\caption{Same as Fig.~\ref{f5} but for the inelastic  \oc scattering to the $2^+_1$
state of $^{12}$C at $E_{\rm lab} = 260$ MeV \cite{Ohk14i}.} \label{f6}
\end{figure}

To explore the intriguing disappearance of Airy structure in inelastic scattering, we employ the extended technique of decomposing the inelastic scattering amplitude into $K$-subamplitudes according to expressions (\ref{eq8}) and (\ref{eq9}). For the case of the $2^+_1$ excitation of $^{12}$C, we have the three subamplitudes groups associated with $K=-2,0,2$ corresponding to $L'=L-2,L'=L,L'=L+2$, respectively. The cross sections of the $K$-subamplitudes obtained for the cases of inelastic scattering at $E_{\rm lab}=124$ and 260 MeV are presented in Figs.~\ref{f5} and \ref{f6}. At the energy $E_{\rm lab}=124$ MeV, with the reduced imaginary potential strength $W_V^{\rm en(ex)}/3$, the prominent Airy minimum A2 of the inelastic scattering cross section can be identified at $\theta_{\rm c.m.}\approx 91.5^\circ$, which is close to the value of $\theta_{\rm c.m.}\approx 88.5^\circ$ for elastic scattering. For the cross sections of the subamplitudes $K=-2,0,2$, because each of these amplitudes has the same $L$ summation scheme as the one in elastic scattering amplitude, we can observe that the cross section of each $K$-subamplitude exhibits a nuclear rainbow structure resembling that of elastic scattering. The prominent A2 minimum in the partial inelastic \oc cross section with $K=0$ can be clearly identified at $\theta_{\rm c.m.}\approx 85^\circ$ (Fig.~\ref{f5}(b)), which is very close to the A2 position of elastic scattering, even without the need to reduce the imaginary part of the OP. Meanwhile, the A2 minimum of the $K=2$ partial inelastic cross section is shifted to smaller angles. The A2 minimum in the $K=-2$ partial inelastic cross section, although less pronounced, can also be identified at larger angles compared to $K=0$. 

To further clarify the formation of the nuclear rainbow in the inelastic scattering to the $2^+_1$ state of $^{12}$C, the total and partial inelastic cross sections with $K=-2,0,2$ are calculated with a less absorptive OP (\ref{eq13b}) $W_V^{\rm en(ex)}\rightarrow W_V^{\rm en(ex)}/3$ (Fig.~\ref{f5}(a)). In this case, we can identify up to the A4 minimum in the partial inelastic cross sections, similar to the one observed in elastic scattering. At large angles, the prominent broad rainbow shoulder A2 of the $K=0$ component is stronger than that of $K=\pm2$, therefore, the nuclear rainbow structure in the inelastic scattering cross section is mainly dominated by this $K=0$ component. The out-of-phase interference of the coupled partial waves with $K=2$ and $K=-2$ smears out the $K=0$ Airy minima, leading to the suppression and smearing of the Airy minimum A2 in the inelastic cross section. Also, because the nuclear rainbow structure of the inelastic scattering is mainly dominated by the $K=0$ component, which is also in phase with the elastic scattering cross section, the Airy minima A2 of the elastic and inelastic scatterings are very close to each other, with only a slight shift of about $\approx 3^\circ$. 


A similar pattern is identified in the CC results for elastic and inelastic scatterings at the other end of the refractive energy range with $E_{\rm lab}$ = 260 MeV (Fig.~\ref{f6}). While a prominent A1 minimum is clearly manifested in the elastic cross section, it vanishes in the inelastic cross section and can only be observed at $\theta_{\rm c.m.}\approx 49^\circ$ with a less absorptive OP $W_V^{\rm en(ex)}/3$. Again, the Airy minimum A1 in the $K=0$ partial scattering cross section is in phase with the elastic scattering one, and the $K=0$ A1 minima can also be faintly seen with a normal absorption as illustrated in the panel (b) of Fig.~\ref{f6}. The rainbow shoulders beyond the Airy minima in the partial inelastic cross sections with $K = \pm2$ are significantly weaker than that with $K = 0$. Similar to the $E_{\rm lab}=124$ MeV case, the out-of-phase interference of the $K=0$ subamplitudes with those of $K=\pm2$ smooth out the Airy minimum A1 structure in the summed inelastic scattering cross section making it visually unobservable in the experimental data. Using the proper $K$ splitting technique for the inelastic scattering amplitude (Eq.~\ref{eq9}), the calculated Airy minima A1 in the elastic and $2^+_1$ inelastic scattering cross sections appear to coincide at $\theta_{\rm c.m.}\approx 49^\circ$.

At energy $E_{\rm lab}$ = 260 MeV, the Airy minima A2 are shifted further towards smaller angles into the Fraunhofer diffractive region. The splitting technique reveals that the magnitude of the partial inelastic cross section with $K$=0 is not significantly dominant compared to $K = \pm2$ ones in this angular region. Therefore, the mixing of the three components $K=-2,0,2$ causes the A2 minimum to become structureless, and the Airy minimum A2 does not present even when reducing the absorptive OP to $W_V^{\rm en(ex)}/3$ (Fig.~\ref{f6}(a)). This observation also holds for the 200 MeV case, as can be seen in Fig.~\ref{f4}. This explains why in the analysis of Ref.~\cite{Ohk17}, only the supernumerary rainbows at energies below 200 MeV can be investigated through the absorption reducing approach.
 
\begin{figure}[bt]\vspace*{-0.4cm}\hspace*{-0.5cm}
	\includegraphics[angle=0,width=0.55\textwidth]{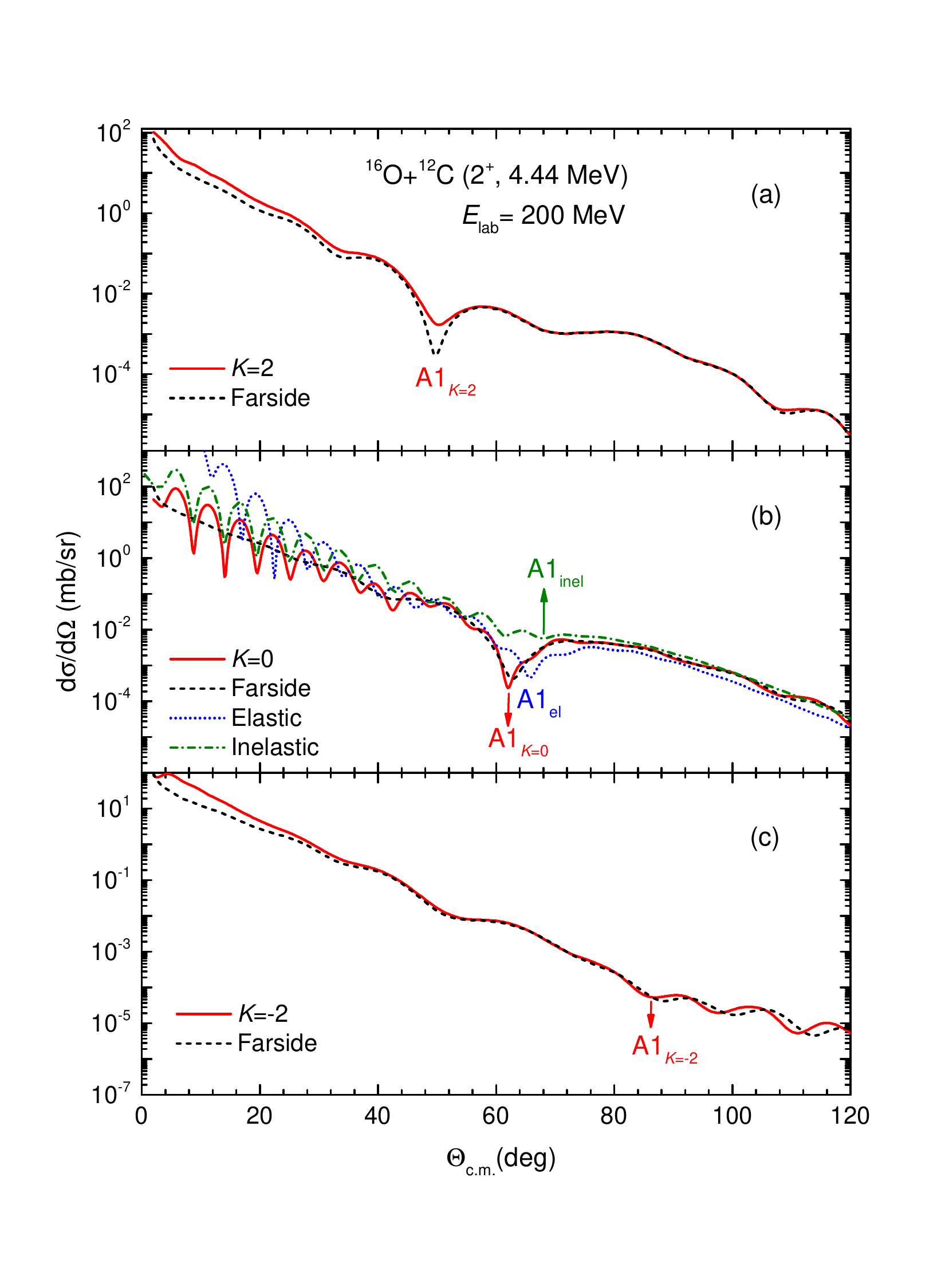}\vspace*{-1.2cm}
	\caption{Farside analysis (dashed lines) of the partial inelastic cross sections with $K=2, 0, -2$ (solid lines). Panel (b) also includes the elastic (dotted line) and inelastic (dashed-dotted line) scattering cross sections.} \label{f7}
\end{figure}

\begin{figure}[bt]\vspace*{0cm}\hspace*{-0.4cm}
	\includegraphics[angle=0,width=0.55\textwidth]{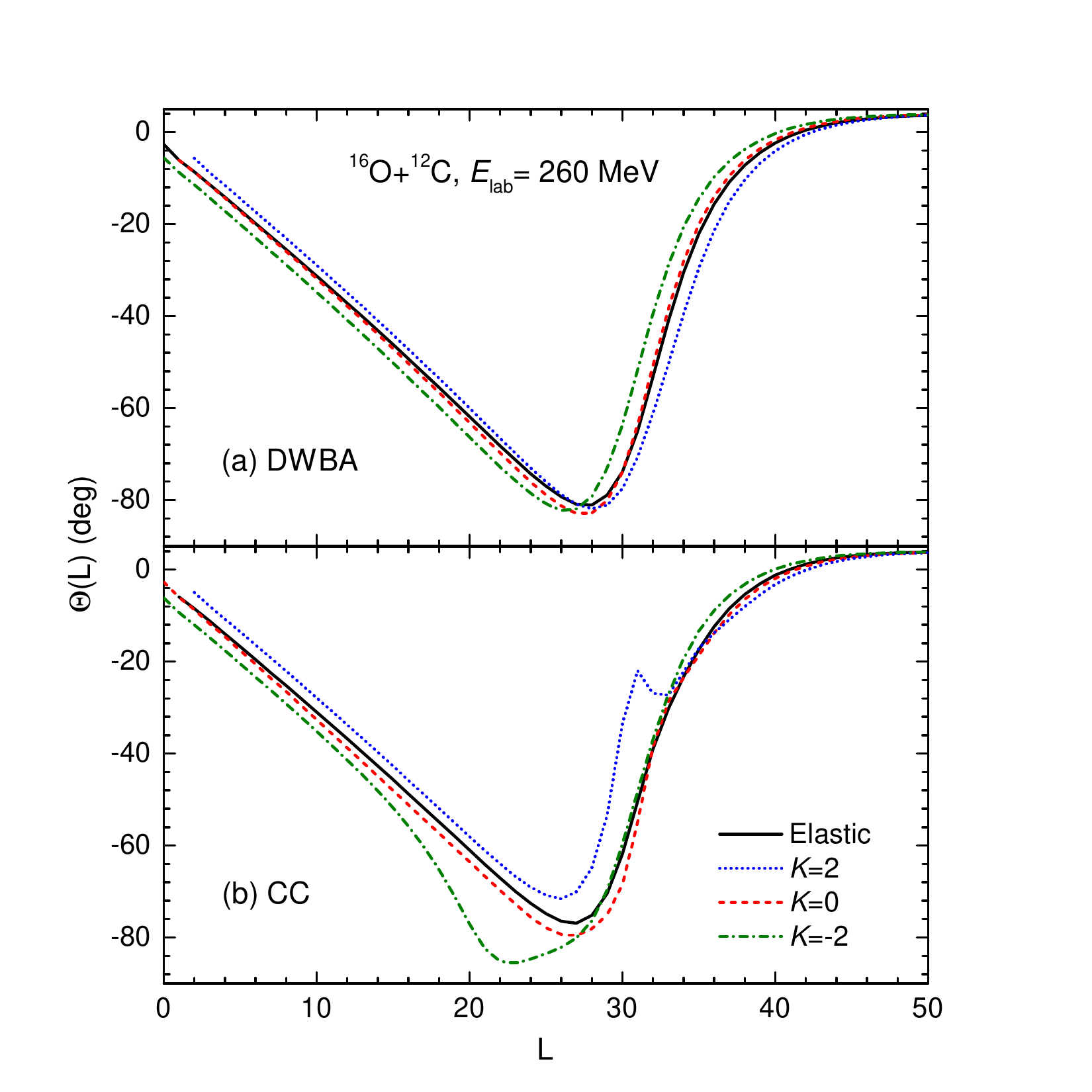}\vspace*{-0.4cm}
	\caption{Deflection functions for elastic (solid lines) and inelastic scattering components corresponding to the three components $K=-2$ (dashed-dotted lines), $K=0$ (dashed lines), and $K=2$ (dotted lines) are provided by the DWBA (a) and CC (b) frameworks, using only the real OP. } \label{f8}
\end{figure}

To elucidate the farside contributions of the subamplitudes with $K=-2,0,2$, we calculated these components at $E_{\rm lab}=200$ MeV, as shown in Fig.~\ref{f7}. The dominance of farside contributions at medium and large angles are evident in the partial inelastic cross sections, with Airy minima A1 identified at $\theta_{\rm c.m.}\approx 49^\circ, 62^\circ$, and $84^\circ$ for $K=2,0,-2$, respectively. The $K=0$ partial inelastic cross section is compared with the elastic and inelastic scatterings cross sections in the middle panel (b). The Airy minimum ${\rm A1}_{K=0}$ is very deep and followed by a broad rainbow shoulder, similar to the one in elastic scattering. This confirms the strong refraction in the partial inelastic scattering cross section and clearly indicates the mixing effect of the $K=2,0,-2$ components as the fundamental cause for the suppression of the Airy structure in the inelastic cross section. As discussed in Ref.~\cite{Phu21}, this phenomenon is analogous to how white light in classical optics results from the mixing of various wavelengths, with the $K$-subamplitudes splitting technique acting as a kind of dispersive lens. 
 
\begin{figure}[bt]\vspace*{0cm}\hspace*{-0.9cm}
	\includegraphics[angle=0,width=0.6\textwidth]{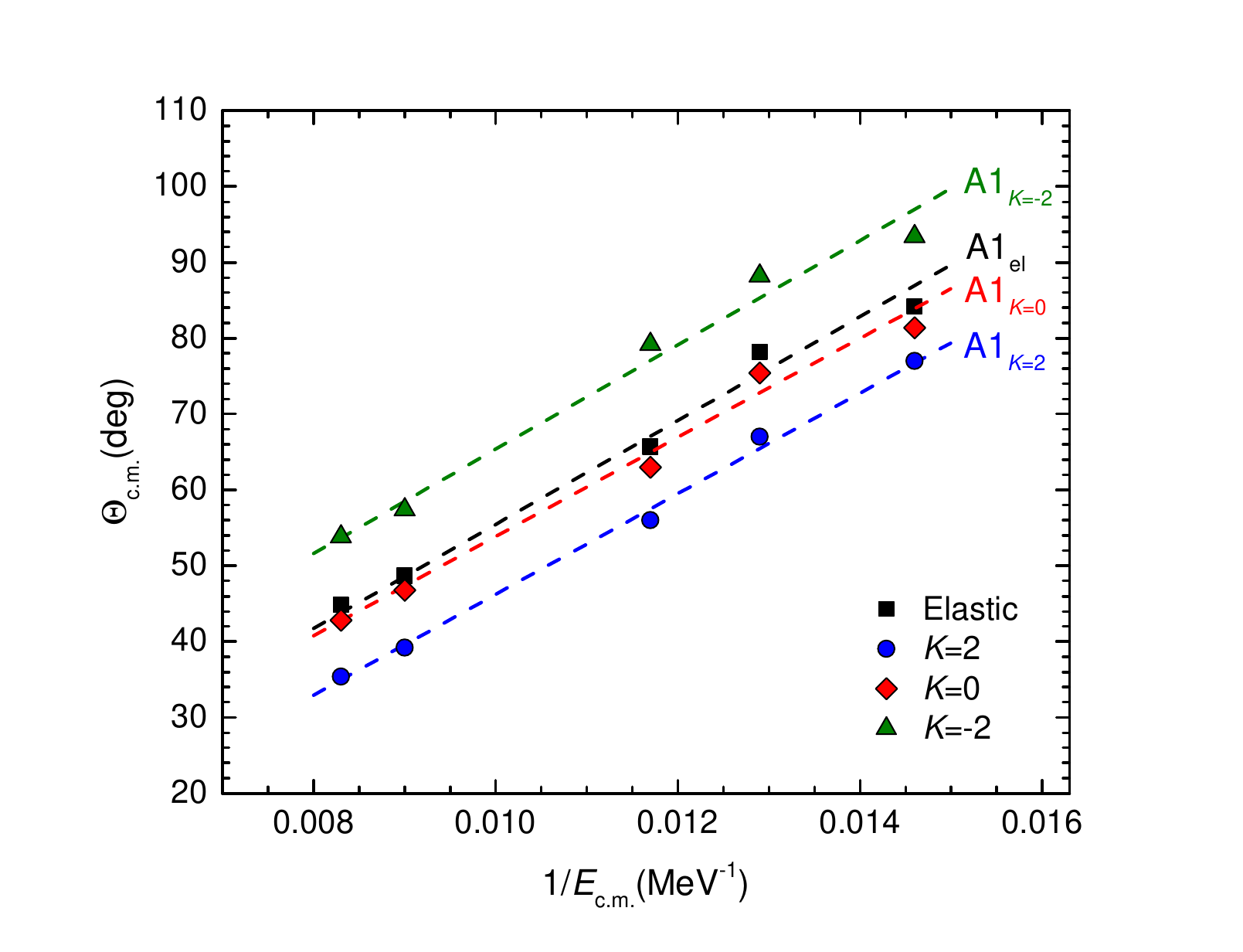}\vspace*{-0.5cm}
	\caption{The positions of the Airy minimum A1 observed in the elastic and partial inelastic scattering cross sections to the $2^+_1$ state of $^{12}$C of \oc are displayed as a function of the inverse center-of-mass (c.m.) energies. A$k$ is the $k$-th order Airy minimum and the lines are to guide the eye.} \label{f9}
\end{figure}  

To investigate the mechanism of the angular-shifted Airy minima in the partial inelastic cross sections, we examined the inelastic deflection functions of the three components $K=-2,0,2$ and compared them with the one of elastic scattering. The deflection functions of the 260 MeV case are presented in Fig.~\ref{f8}. The CC calculation results (panel (b)) show that for each entrance orbital angular momentum $L<L_{\rm R}$ ($L_{\rm R}\sim$ 25--28), the partial deflection function $\Theta^K(L)$ follows the rule $\lvert \Theta^{K=2}\rvert<\lvert \Theta^{K=0}\rvert<\lvert \Theta^{K=-2}\rvert $. We know that the $L$ values corresponding to $L< L_{\rm R}$ contribute mainly to the farside $f^{\rm F_<}$ component at medium and large angles \cite{Phu24,An01,McV92}. According to the semi-classical approximation, there is a relationship between the deflection function and the scattering angle $\theta_{\rm c.m.}$=$\lvert\Theta^K(L)\rvert$ \cite{Fro83,Hus84}. Therefore, the larger the deflection angle, the more the contribution of $f^{\rm F_<}$ shifts towards larger angles. Consequently, the nuclear rainbow structure is also likely to shift towards larger angles. This explains the shift of Airy minima A$k$ corresponding to the three components $K=-2,0,2$ in the partial inelastic scattering cross sections in the order ${\rm A}k_{K=2}<{\rm A}k_{K=0}<{\rm A}k_{K=-2}$.

For the case of $K=0$, the deflection function corresponding to the $K=0$ partial inelastic scattering is considerably similar to the elastic scattering deflection function, resulting in an Airy structure in the $K=0$ partial inelastic cross section that tends to be in phase with the elastic scattering cross section. As a consequence, Fig.~\ref{f9} shows that the evolution of Airy minimum A1 for the $K=0$ component is quite close to the Airy minimum of elastic scattering. The phase shift in the deflection functions of the three components $K=-2,0,2$ also leads to the separation into three Airy minima A1 evolution paths, as shown in Fig.~\ref{f9}. It should be noted that for the $K=2$ case, the deflection function has a rather irregular structure with a local minimum at $L \approx 32$. This is due to coupling effects in CC calculations as the DWBA results (without coupling effects) in panel (a) do not exhibit this irregular structure.
 
\begin{figure}[bt]\vspace*{0cm}\hspace*{-1.2cm}
	\includegraphics[angle=0,width=0.6\textwidth]{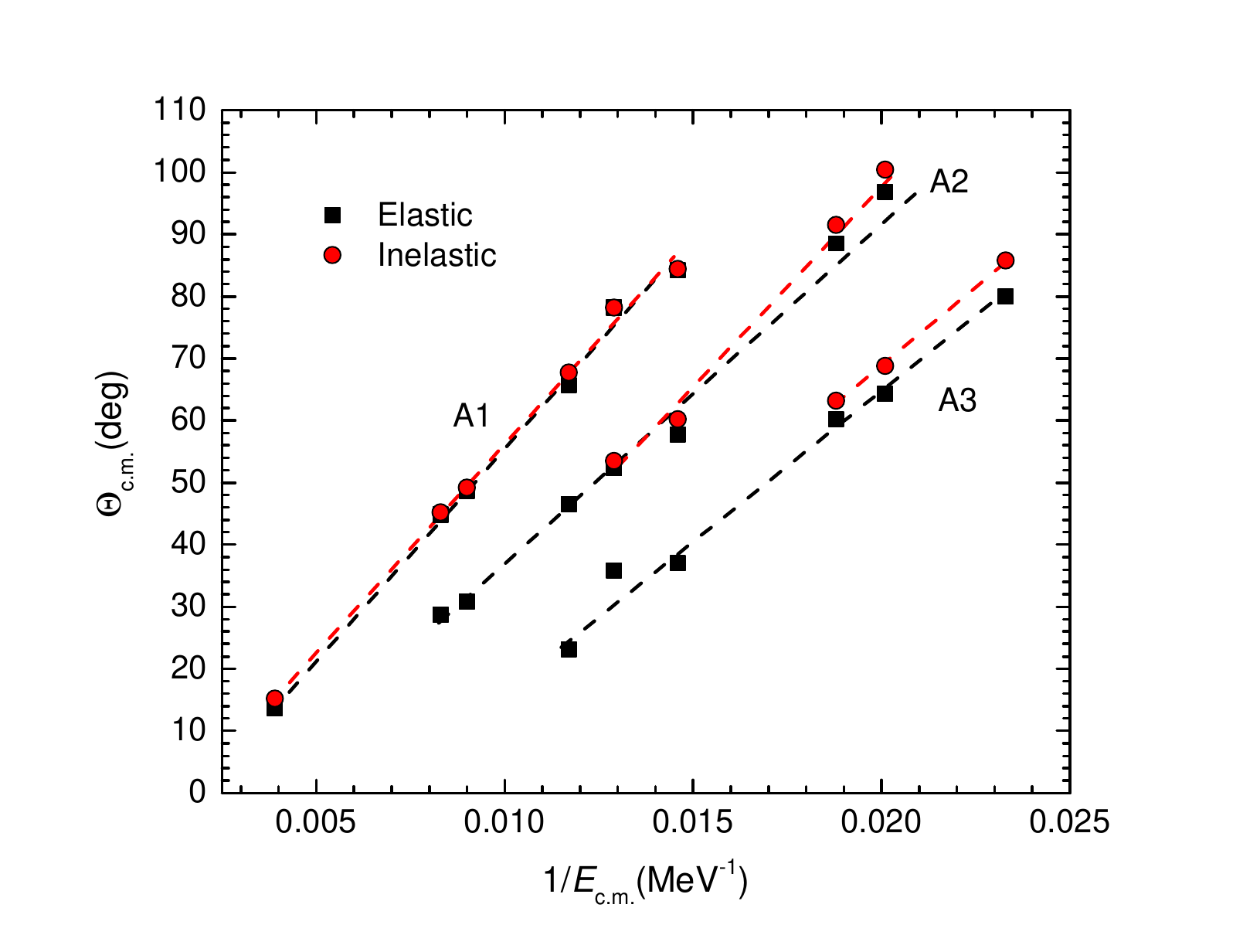}\vspace*{-0.5cm}
	\caption{The positions of the Airy minimum observed in the elastic (squares) and inelastic scattering to the $2^+_1$ state of $^{12}$C (circles) of \oc are displayed as a function of the inverse center-of-mass (c.m.) energies. A$k$ is the $k$-th order Airy minimum and the lines are to guide the eye.} \label{f10}
\end{figure}

Finally, the results describing the systematic evolution of Airy minima of elastic and inelastic \oc scatterings, presented in Fig.~\ref{f10}, consistently indicate that A1 minima of elastic and inelastic scatterings are quite close to each other in the refractive energy range from 170 to 608 MeV. For A2, at energies $E_{\rm lab}\geq 170$ MeV, the Airy minima A2 of elastic and inelastic scatterings are very close, similar to the A1 case. However, at lower energies $E_{\rm lab}=124$ and 115.9 MeV, there is a slight shift of $\approx 3^\circ-4^\circ$ towards larger angles of Airy minima A2 in inelastic scattering compared to the elastic one. The shift further increases at lower energies. A similar trend is also observed for Airy minima A3. At the lowest considered energy of $E_{\rm lab}=100$ MeV, the shift of Airy minima A3 in inelastic scattering is about $6^\circ$.

In a previous study by Michel et al. \cite{Mi04}, it was pointed out that the shift of Airy minima in inelastic scattering cross sections compared to elastic scattering is due to the loss of the kinetic energy of $E_{\rm cm}-E_{x}(2^+_1)$ in the excited channel. For the \oc scattering system, the primary rainbow with A1 minima only appears at medium energies corresponding to $E_{\rm lab} >170$ MeV ($E_{\rm cm}\approx 73$ MeV). At these energies, the change in $E_{\rm cm}-E_{x}(2^+_1)$ in the excited channel compared to the elastic scattering channel is negligible, resulting in small shifts of Airy minima A1 and A2 in this energy range. However, at lower energies corresponding to $E_{\rm lab} <124$ MeV ($E_{\rm cm}\approx 53$ MeV), the change in $E_{\rm cm}-E_{x}(2^+_1)$ in the excited channel becomes increasingly considerable compared to the elastic scattering channel. As a result, the shift of Airy minima A2 and A3 increases as the energy decreases. We note that this observation is also consistent with the analysis by Michel et al. \cite{Mi04} on the $\alpha$+$^{40}$Ca scattering to the $3^-$ state ($E_{x}(3^-)=3.73$ MeV) of $^{40}$Ca over a wide energy range of $E_{\rm lab}$ from 29 MeV to 100 MeV ($E_{\rm cm}\approx$ 26--91 MeV).

It should be emphasized that the similarity in positions of the Airy minima A1 of inelastic and elastic cross sections in our CC calculations differs from the previous conclusion of Ref.~\cite{Ohk17}, which suggested that A1 minima in the inelastic scattering cross sections are consistently shifted towards larger angles compared to those in the elastic scattering cross sections (see also Fig.~5 of the mentioned work). However, careful examination of the CC calculation results shown in Fig.~2 of Ref.\cite{Ohk17} reveals that the Airy minima A1 of elastic and inelastic scattering are very close to each other for most of the energies. This discrepancy between the CC calculations in Fig.~2 and the discussion in Fig.~5 of Ref.\cite{Ohk17} arises because in Fig.~5 the authors identified the Airy minima of elastic scattering based on experimental data while the minima of inelastic scattering were based on the CC calculation results. Meanwhile, most of the CC calculation results in Ref.~\cite{Ohk17} show Airy minima A1 in elastic scattering cross sections shifted towards larger angles compared to experimental elastic scattering data, which may also lead to a shift of Airy minima A1 in inelastic scattering cross sections towards larger angles. For elastic scattering, identifying Airy minima visually is generally quite unambiguous due to the deep shape of the minima. However, for inelastic scattering leading to the $2^+_1$ state, as discussed above, Airy minima of inelastic scattering are suppressed and broadened, leading to significant uncertainty in identifying inelastic Airy minima. This results in CC calculations with Airy minima A1 in inelastic scattering cross sections, although shifted towards larger angles, appearing to agree with experimental data. This is the fundamental reason behind the reported shift in Airy minima of the inelastic scattering cross section in Ref.\cite{Ohk17}.

\section{Summary}

Nuclear rainbows in elastic and inelastic \oc scattering exciting the $2^+_1$ state of $^{12}$C have been systematically analyzed for $E_{\rm lab}$ ranging from 100 to 608 MeV. By employing a novel $K$-subamplitudes splitting technique combined with an extension of Fuller's NF decomposition for the inelastic scattering amplitude, we have elucidated the formation of nuclear rainbows for primary and supernumerary bows in inelastic \oc scattering over a wide range of refractive energies. For excitation to the $2^+_1$ state of $^{12}$C, the inelastic scattering amplitude comprises a superposition of three subamplitudes with $K=2,0,-2$, each having the similar Airy minima structure with the others and the one of the elastic scattering amplitude. However, these Airy minima structures for $K=2,0,-2$ amplitudes are out of phase with each other. Consequently, the mixing of these three components suppresses the Airy minima structure of the inelastic scattering cross section. This explains why the Airy minima of inelastic scattering are shallow and difficult to observe experimentally compared to elastic scattering. We also observed that at medium and large angles, the Airy structure from the $K=0$ component is stronger than those of $K = \pm2$. Therefore, the nuclear rainbow structure in the total inelastic cross section exciting the 2$^+$ state is primarily governed by this $K=0$ component. Since the partial inelastic \oc cross section with $K=0$ has an Airy structure in phase with elastic scattering, this explains the considerable similarity between the positions of the Airy minima of inelastic and elastic scatterings. 

To further explore the mechanism of rainbow formation, we also propose, for the first time, the partial deflection functions for each $K$-subamplitude of inelastic scattering for interpreting the angular-shifted Airy minima structure of the partial inelastic cross sections for $K=2,0,-2$. The partial deflection functions follow the order $\lvert\Theta^{K=2}\rvert<\lvert\Theta^{K=0}\rvert<\lvert\Theta^{K=-2}\rvert$ in the region $L<L_{\rm R}$, which explains the order of the Airy minima positions in the $K$-subamplitudes. Furthermore, the partial deflection function of the $K=0$ component closely resembles that of elastic scattering, resulting in the Airy structure of the $K=0$ component being in phase with elastic scattering. 
 
The combined use of the $K$-subamplitudes splitting technique in the inelastic scattering amplitude, NF decomposition, and deflection function allows us to unambiguously determine the systematics of nuclear rainbow Airy minima for inelastic \oc scattering to the $2^+_1$ state of $^{12}$C. At higher energies than 200 MeV, we cannot observe the secondary Airy minimum A2 even with reduced imaginary potential. This is because, at these energies, Airy minima are shifted into the diffractive Fraunhofer region at small angles. Additionally, the Airy minimum A2 amplitude with $K=0$ is not significantly dominant compared to the other two components $K=\pm2$ in this region. Therefore, the mixing of the three components $K=-2,0,2$ destroys the A2 minimum structure in the total amplitude, making it impossible to identify this structure.

In the medium energy range with 170 MeV $\leq E_{\rm lab} \leq 200$ MeV, the A1 and A2 Airy minima are clearly revealed. Since the system has lost negligible kinetic energy in the inelastic scattering channel compared to the elastic one, there is no considerable shift in the A1 and A2 minima and their positions in the inelastic scattering cross sections almost coincide with those in the elastic ones. At lower energies $E_{\rm lab}=100$--$124$ MeV, where only secondary Airy minima A2 and A3 can be identified in the inelastic scattering cross sections, the shift of Airy minima becomes more pronounced, ranging from $\approx 3^\circ-6^\circ$, due to the notable difference in the kinetic energy lost between the two channels. Ultimately, our analysis shows that the shifts of Airy A2 and A3 minima between the elastic and inelastic \oc scatterings increase as $E_{\rm lab}$ decreases while there is no sizable shift for the A1 minima. 

These systematic analyses demonstrate the effectiveness of the combined use of $K$-subamplitudes splitting technique in the inelastic amplitude, NF decomposition, and deflection function in understanding the formation of nuclear rainbows in the inelastic scattering of nuclear systems. These findings using the methods will open new avenues for exploring other light heavy-ion and alpha-nucleus systems. Furthermore, the applications of these techniques to atomic and molecular collision research, where quantum rainbow scattering also manifests, may yield valuable insights. 

\section*{Acknowledgments}
We thank Dao Tien Khoa for his enlighten discussions. The present research has been supported, 
in part, by the National Foundation for Science and Technology Development 
(NAFOSTED Project No. 103.04-2021.74). 

\bibliographystyle{apsrev4-2}
\bibliography{references}
\end{document}